\documentclass[doublecol]{epl2} 
\usepackage{graphicx}
\usepackage{dcolumn}
\usepackage{bm}
\usepackage{amsmath}
\usepackage{mathrsfs}
\usepackage{subfigure}

\title{Percolation on a maximally disassortative network}

\author{Shogo Mizutaka\thanks{\email{shogo.mizutaka.sci@vc.ibaraki.ac.jp}} \and Takehisa Hasegawa\thanks{\email{takehisa.hasegawa.sci@vc.ibaraki.ac.jp}}}
\shortauthor{S. Mizutaka \& T. Hasegwa}

\institute{                  
  Department of Mathematics and Informatics, Ibaraki University, 2-1-1 Bunkyo, Mito 310-8512, Japan\\
}
\pacs{64.60.aq}{Networks}
\pacs{64.60.ah}{Percolation}
\pacs{68.35.Rh}{Phase transitions and critical phenomena}

\abstract{
We propose a maximally disassortative (MD) network model which realizes a maximally negative degree-degree correlation, and study its percolation transition to discuss the effect of a strong degree-degree correlation on the percolation critical behaviors.
Using the generating function method for bipartite networks, we analytically derive the percolation threshold and the order parameter critical exponent, $\beta$. For the MD scale-free networks, whose degree distribution is $P(k) \sim k^{-\gamma}$, we show that the exponent, $\beta$, for the MD networks and corresponding uncorrelated networks are same for $\gamma>3$ but are different for $2<\gamma<3$.
A strong degree-degree correlation significantly affects the percolation critical behavior in heavy-tailed scale-free networks.
Our analytical results for the critical exponents are numerically confirmed by a finite-size scaling argument.
}

\begin{document}

\maketitle
\section{Introduction}
Numerous complex systems are abstracted as networks consisting of simplified elements (nodes) and their connections (edges) e.g., the Internet, World Wide Web, and prey/predator relations in ecosystems \cite{Caldarelli_text}.
It is well known that most real-world networks are scale-free, such that the degree distribution $P(k)$ obeys the power-law function: $P(k)\sim k^{-\gamma}$, where $\gamma$ is called the degree exponent.
The robustness of networks to failures and attacks has been frequently discussed by considering the network percolation problem.
Some studies concerning percolation models on networks have shown that scale-free networks are extremely robust to the random removal of nodes and edges; however, they are fragile to the targeted removal of the high-degree nodes \cite{Newman_text}.

Percolation on networks has been studied in order to also understand the relation between the critical phenomena and underlying network structures.
In a seminal work concerning the critical behavior of percolation transitions on (degree-)uncorrelated scale-free networks, Cohen {\it et al.} \cite{Cohen02} analytically derived the relation between the percolation critical exponents and degree exponent, $\gamma$.
 Specifically, an unconventional universality class emerges for $2<\gamma <4$, and a mean-field class is observed for $\gamma \ge 4$.
In addition, the fractal dimension of a percolating cluster in a critical state \cite{Cohen04} and the upper critical dimension for uncorrelated scale-free networks \cite{Wu07} have been established.

In real-world networks, a degree-degree correlation, which is the correlation of the degrees of the nodes directly connected by an edge, would arise \cite{Newman02,Newman03}.
In a network with a positive (negative) degree-degree correlation, similar (dissimilar) degree nodes tend to connect to each other.
A degree-degree correlated structure can affect critical phenomena on a network.
Goltsev {\it et al.} \cite{Goltsev08} treated analytically percolation on degree-degree correlated networks  using the eigenvectors and associated eigenvalues of a branching matrix defined by the conditional probability, $P(k|k')$, that a random neighbor of a degree-$k'$ node has degree $k$. 
They showed the necessary and sufficient conditions that the critical behavior of a percolation on a degree-degree correlated network is the same as that on uncorrelated networks with an identical degree distribution.
When a network does not satisfy any of their conditions, its critical behavior does not coincide with that of the corresponding uncorrelated networks.
Two strongly correlated networks analyzed in \cite{Goltsev08} violate one of their conditions, and so, exhibit an atypical universality class depending on the details of the network structure.
Further studies are needed to attain a better understanding of the critical behavior of correlated networks; however, 
there is little research on the investigation of percolation transition on strongly correlated networks owing to the lack of other solvable models.

In this study, we propose a solvable model in which the networks realize a maximally negative degree-degree correlation. Hereinafter, we call them maximally disassortative (MD) networks.
Applying the generating function method for bipartite networks to the percolation on the MD networks, we analytically derive the percolation threshold and the order parameter critical exponent, $\beta$, related to the relative size of the giant component.
For the MD scale-free networks with $2<\gamma<3$, an unconventional critical behavior is observed: the critical exponent, $\beta$, acquires a value that is different from that of the uncorrelated networks.
Contrastingly, the MD scale-free networks with $\gamma\ge 3$ belong to the same universality class as that of the uncorrelated ones.
A strong degree-degree correlation significantly affects the critical behavior in heavy-tailed scale-free networks.
Our analytical estimations are confirmed by a finite-size scaling analysis near the zero percolation threshold \cite{Radicchi15}.

%%%%%%%%%%%%%%%%%%%%%%%
\begin{figure}[t]
\onefigure[width=0.45\textwidth]{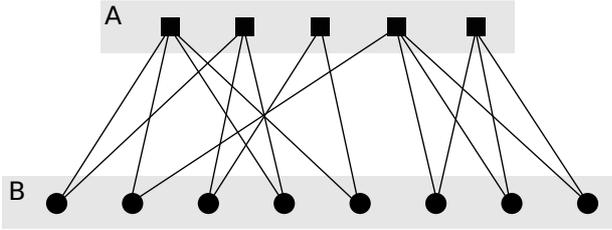}
\caption{Illustration of a bipartite network that realizes an MD network. The degree of group A nodes (squares) is distributed, whereas that of group B nodes (circles) is two.}
\label{fig:zrrg}
\end{figure}
%%%%%%%%%%%%%%%%%%%%%%%

%%%%%%%%%%%%%%%%%%%%%%%%%%%%%%%%%%
%%%%%%%%%%%%%%%%%%%%%%%%%%%%%%%%%%
\section{Maximally disassortative network}\label{sec:MDN}
Let us construct our MD networks as bipartite networks in which each node belongs to either of the two groups, A or B, and each edge connects a group A node and a group B node (Fig.~\ref{fig:zrrg}).
First, the number, $N_{\rm A}$, of nodes in group A is given, and the number of stubs, i.e., degree of each node in group A is assigned by a predetermined degree distribution, $P_{\rm A}(k)$.
In this study, we consider a power-law degree distribution for group A, i.e.,
\begin{eqnarray}
	P_{\rm A}(k) \sim k^{-\gamma},
\end{eqnarray}
for large $k$.
Next, to realize a maximally negative degree-degree correlation, we designate a minimum degree, $k_{\rm min}$, of the network, prepare $N_{\rm B}$ nodes in group B, and let all the group B nodes have degree $k_{\rm min}$. 
Thus, the degree distribution, $P_{\rm B}(k)$, for group B is
\begin{eqnarray}
	P_{\rm B}(k)=\delta_{k,k_{\rm min}},
\end{eqnarray}
where $\delta$ is the Kronecker delta.
For a network to be bipartite, the total number of degrees in group A should be equal to that in group B.
Then, the number, $N_{\rm B}$, of nodes in group B is determined from the relation, $N_{\rm A}\sum_{k}kP_{\rm A}(k)=N_{\rm B}\sum_{k}kP_{\rm B}(k)$, or equivalently,
\begin{eqnarray}
	z_{1}^{\rm A}N_{\rm A}=z_{1}^{\rm B}N_{\rm B},
\end{eqnarray}
where $z_{1}^{\rm A}=\sum_{k}kP_{\rm A}(k)$ ($z_{1}^{\rm B}=k_{\rm min}$) is the average degree of group A (B).
An edge is formed by randomly selecting a stub from each of groups and joining them. This process is repeated until no stub exists. Then, a network with the degree distribution,
\begin{eqnarray}
	 P(k)=rP_{\rm A}(k)+(1-r)P_{\rm B}(k),
	 \label{eq:degree_distribution}
\end{eqnarray}
is realized. Here, $r=N_{\rm A}/(N_{\rm A}+N_{\rm B})$.
Because any node in group A is connected to only the group B nodes having the minimum degree, $k_{\rm min}$, the degree-degree correlation of the entire network is totally negative, i.e., disassortative.
This type of network realizes an MD structure in that any edge swapping increases the Spearman's rank correlation coefficient for the network \cite{Fujiki17} \footnote{In \cite{Fujiki17}, the degree correlation of a hierarchical scale-free network called $(u,v)$-flower \cite{Rozenfeld07}  was discussed. Spearman's rank correlation coefficient was utilized to evaluate the negative degree correlation correctly because Pearson's correlation coefficient does not work for disassortative networks with the degree exponent $\gamma<4$ \cite{Litvak13}.}.

%%%%%%%%%%%%%%%%%%%%%%%%%%%%%%%%%%
%%%%%%%%%%%%%%%%%%%%%%%%%%%%%%%%%%
\section{Percolation on bipartite networks}\label{sec:percolation}
We briefly recall the generating function method for percolation on bipartite networks with arbitrary $P_{\rm A}(k)$ and $P_{\rm B}(k)$ \cite{Newman02-2} prior to the analysis of the MD networks.
We introduce the generating functions for the degree distributions of groups A and B, i.e.,
\begin{eqnarray}
&&	G^{\rm A}_{0}(x)= \sum_{k=0}P_{\rm A}(k)x^{k}, \\
&&	G^{\rm B}_{0}(x)= \sum_{k=0}P_{\rm B}(k)x^{k}.	
\end{eqnarray}
Similarly, the generating functions, $G^{\rm A}_{1}(x)$ and $G^{\rm B}_{1}(x)$, for the so-called excess degree distributions of groups A and B are given by
\begin{eqnarray}
	G^{\rm A}_{1}(x)&& = \sum_{k=0}\frac{kP_{\rm A}(k)}{{z}_{1}^{\rm A}}x^{k-1}, \label{eq:G1A}\\
	G^{\rm B}_{1}(x)&& = \sum_{k=0}\frac{kP_{\rm B}(k)}{z_{1}^{\rm B}}x^{k-1},\label{eq:G1B}
\end{eqnarray}	
respectively.

Let us consider the site percolation process on a bipartite network. 
Each node is occupied with probability $p$ and unoccupied with probability $q=1-p$.
Let $H_{1}^{\rm A}(x)$ [$H_{1}^{\rm B}(x)$] be the generating function for the probability of reaching a branch of a finite size by an edge outgoing from a node in group B (A). 
Under the assumption that a given network is locally tree-like, $H_{1}^{\rm A}(x)$ and $H_{1}^{\rm B}(x)$ satisfy the following equations:
\begin{eqnarray}
&&	H_{1}^{\rm A}(x)= q +pxG_{1}^{\rm A}\left[H_{1}^{\rm B}(x)\right], 	\label{eq:H1A}\\
&&	H_{1}^{\rm B}(x)= q +pxG_{1}^{\rm B}\left[H_{1}^{\rm A}(x)\right].
	\label{eq:H1B}
\end{eqnarray}
Substituting $x=1$ for Eqs.~(\ref{eq:H1A}) and (\ref{eq:H1B}), we have
\begin{eqnarray}
&&	u= q +pG_{1}^{\rm A}(v) \label{eq:u}, \\
&&	v= q +pG_{1}^{\rm B}(u) \label{eq:v},
\end{eqnarray}
where $u=H_{1}^{\rm A}(1)$ and $v=H_{1}^{\rm B}(1)$ are the probabilities to reach a finite branch by an edge outgoing from a group B node and a group A node, respectively.
In the percolating phase, i.e., $p>p_{\rm c}$, $u< 1$ and $v< 1$.

To derive the percolation threshold for the bipartite network, we introduce the generating function, $H_{0}^{\rm tot}(x)$, for the probability of a node belonging to a cluster of a finite size:
\begin{eqnarray}
	H_{0}^{\rm tot}(x)=rH_{0}^{\rm A}(x)+(1-r)H_{0}^{\rm B}(x),
	\label{eq:H0tot}	
\end{eqnarray}
where $H_{0}^{\rm A}(x)$ [$H_{0}^{\rm B}(x)$] is the generating function for the probability of a node in group A (B) belonging to a cluster of a finite size as
\begin{eqnarray}
	H_{0}^{\rm A}(x) &=q+pxG_{0}^{\rm A}\left[H_{1}^{\rm B}(x)\right], \label{eq:H0A} \\
	H_{0}^{\rm B}(x) &=q+pxG_{0}^{\rm B}\left[H_{1}^{\rm A}(x)\right]\label{eq:H0B}.	
\end{eqnarray}
Thus, the size of the giant component, $S$, is given by 
\begin{eqnarray}
	S&=&1-H_{0}^{\rm tot}(1) \nonumber \\
	&=& p\left(1-rG_{0}^{\rm A}(v)-(1-r)G_{0}^{\rm B}(u)\right)\label{eq:GCsize},
\end{eqnarray}
and the average size of the finite clusters, $\langle s\rangle$, is given as $\langle s\rangle=dH_{0}^{\rm tot}(x)/dx|_{x=1}$. 
In the non-percolating phase ($p<p_{\rm c}$) where $H_{1}^{\rm A}(1)=H_{1}^{\rm B}(1)=1$, this average cluster size reduces to
\begin{eqnarray}
	\langle s\rangle=p+prG_{0}^{\rm A'}(1)H_{1}^{\rm B'}(1)+p(1-r)G_{0}^{\rm B'}(1)H_{1}^{\rm A'}(1),~~~
	\label{eq:av_s}
\end{eqnarray}
where $H_{1}^{\rm A'}(1)$ and $H_{1}^{\rm B'}(1)$ are 
\begin{eqnarray}
	H_{1}^{\rm A'}(1)
	=\frac{p+p^2 z_{2}^{\rm A}/z_{1}^{\rm A} }{1-p^{2}z_{2}^{\rm A}z_{2}^{\rm B}/(z_{1}^{\rm A}z_{1}^{\rm B})}
	\label{eq:H1A'}
\end{eqnarray}
and
\begin{eqnarray}
	H_{1}^{\rm B'}(1)=\frac{p+p^2 z_{2}^{\rm B}/z_{1}^{\rm B} }{1-p^{2}z_{2}^{\rm A}z_{2}^{\rm B}/(z_{1}^{\rm A}z_{1}^{\rm B})},
	\label{eq:H1B'}
\end{eqnarray}
respectively. Here, $z_{2}^{\rm X}=\sum_{k}k(k-1)P_{\rm X}(k)$.
The percolation threshold is given as the point at which $\langle s\rangle$ diverges.
Equations (\ref{eq:av_s})--(\ref{eq:H1B'}) lead to the result that $\langle s\rangle$ diverges at $p=p_{\rm c}$ given by
\begin{eqnarray}
	p_{\rm c}=\sqrt{\dfrac{z_{1}^{\rm A}}{z_{2}^{\rm A}}\dfrac{z_{1}^{\rm B}}{z_{2}^{\rm B}}}.
	\label{eq:pc}
\end{eqnarray}
The percolation threshold (\ref{eq:pc}) corresponds to that for bond percolation on bipartite networks, which has been previously obtained by several approaches \cite{Newman02-2,Allard09,Hooyberghs10,Bianconi17}.

%%%%%%%%%%%%%%%%%%%%%%%%%%%%%%%%%%
%%%%%%%%%%%%%%%%%%%%%%%%%%%%%%%%%%
\section{Criticality of percolation on MD network}\label{sec:theory}

To discuss the effect of the MD structures on critical behavior, we concentrate on the MD networks having $P_{\rm A}(k)\sim k^{-\gamma}$ ($k\ge 2$) and $P_{\rm B}(k)=\delta_{k2}$.
Applying Eq.~(\ref{eq:pc}) to the MD networks, we obtain the percolation threshold  as
\begin{eqnarray}
p_{\rm c}=\sqrt{\frac{z_{1}^{\rm A}}{z_{2}^{\rm A}}}.
\label{eq:pc_MDN}
\end{eqnarray}
Because Eq.~(\ref{eq:G1B}) reduces to $G_{1}^{\rm B}(x)=x$ in the present case, we have
\begin{eqnarray}
	u&=& q +pG_{1}^{\rm A}\left(q +pu\right)\nonumber \\
	 &=& 1-p +\frac{p}{z_{1}^{\rm A}}\sum_{k}kP_{\rm A}(k)\left(1-p(1-u)\right)^{k-1} \label{eq:u2}
\end{eqnarray}
from Eqs.~(\ref{eq:u}) and (\ref{eq:v}).
At $p=p_{\rm c}+\delta$, where $\delta$ is a positive infinitesimal value, $u$ is slightly smaller than unity, i.e., $u=1-\epsilon$. 
Here, $\epsilon$ is the order parameter and a positive infinitesimal value.
From Eq.~(\ref{eq:u2}), we have
\begin{eqnarray}
	\epsilon= p_{\rm c}+\delta -\frac{(p_{\rm c}+\delta)}{z_{1}^{\rm A}}\sum_{k}kP_{\rm A}(k)\left(1-p\epsilon\right)^{k-1} \label{eq:ep},
\end{eqnarray}
at $p=p_{\rm c}+\delta$.
The summation in Eq.~(\ref{eq:ep}) determines the critical behavior of the percolation on the MD networks.
For $P_{\rm A}(k)\sim k^{-\gamma}$ with $\gamma>3$ in which $p_{\rm c}>0$, the summation in Eq.~(\ref{eq:ep}) has an asymptotic form (See appendix),
\begin{eqnarray}
	&&\sum_{k}kP_{\rm A}(k)(1-p\epsilon)^{k-1}\nonumber \\
	&&~~\sim   z_{1}^{\rm A}- z_{2}^{\rm A}p\epsilon+\frac{1}{2}z_{3}^{\rm A}(p\epsilon)^2 + \cdots+ C(p\epsilon)^{\gamma-2},
	\label{eq:expansion}	
\end{eqnarray}
where the highest order of the analytic term is the largest integer less than $\gamma-2$, $z_{3}^{\rm A}=\sum_{k}k(k-1)(k-2)P_{\rm A}(k)$ and $C$ is a constant.
Substituting Eq.~(\ref{eq:expansion}) into Eq.~(\ref{eq:ep}),
we obtain
\begin{eqnarray}
	\frac{2}{p_{\rm c}}\delta=\frac{1}{2}\frac{z_{2}^{\rm A}}{z_{1}^{\rm A}}p^3\epsilon+\cdots+\frac{C}{z_{1}^{\rm A}}p^{\gamma-1}\epsilon^{\gamma-3}.
	\label{eq:delta-epsilon1}
\end{eqnarray}
Because for $3<\gamma<4$, the leading term of the right-hand side in Eq.~(\ref{eq:delta-epsilon1}) is $Cp^{\gamma-1}\epsilon^{\gamma-3}/z_{1}^{\rm A}$, we approximate $p$ by $p_{\rm c}$ and obtain
\begin{eqnarray}
	\delta\sim
	\frac{C}{2 z^{\rm A}_{1}}p_{\rm c}^{\gamma}\epsilon^{\gamma-3}.
\end{eqnarray}
Consequently, the order parameter, $\epsilon$, is related to the difference, $\delta=p-p_{\rm c}$, as
\begin{eqnarray}
	\epsilon\sim (p-p_{\rm c})^{\frac{1}{\gamma-3}},
\end{eqnarray}
which implies that the critical exponent, $\beta$, related to the relative size, $S$, of the giant component, $S \sim \delta^\beta$, is $\beta_{\rm MDN}=1/(\gamma-3)$ for the MD networks.
This value corresponds to $\beta_{\rm UCN}$ for the uncorrelated scale-free networks having the same degree exponent $\gamma$ \cite{Cohen02,Goltsev08}.
For $\gamma>4$, we easily find the mean-field result:
\begin{eqnarray}
	\epsilon\sim (p-p_{\rm c}),
\end{eqnarray}
i.e., $\beta_{\rm MDN}=\beta_{\rm UCN}=1$.

When $2<\gamma<3$, the percolation threshold is $p_{\rm c}=0$, and the summation in Eq.~(\ref{eq:ep}) becomes
\begin{eqnarray}
	\sum_{k}kP_{\rm A}(k)(1-p\epsilon)^{k-1}\sim z^{\rm A}_{1}+C(p\epsilon)^{\gamma-2}.
	 \label{eq:exp23}
\end{eqnarray}
Substituting Eq.~(\ref{eq:exp23}) and $p=\delta$ into Eq.~(\ref{eq:ep}), we derive
\begin{eqnarray}
	1&-&\epsilon\sim 1-\delta+\frac{\delta}{z_{1}^{\rm A}}\left(z_{1}^{\rm A}+C(\delta\epsilon)^{\gamma-2}\right) \nonumber\\
	&&\iff\epsilon\sim\left(\frac{-C}{z_{1}^{\rm A}}\right)\delta^{\frac{\gamma-1}{3-\gamma}}.
\end{eqnarray}
Thus, for $2<\gamma<3$, the order parameter and the size of the giant component near $p_{\rm c}=0$ behave as
\begin{eqnarray}
	\epsilon & \sim & p^{\beta_{\rm MDN}},\\
	\label{eq:beta23}
	S & \sim & p^{\beta_{\rm MDN}+1},
	\label{eq:S23}
\end{eqnarray}
where $\beta_{\rm MDN}=(\gamma-1)/(3-\gamma)$. 
For $2<\gamma<3$, $\beta_{\rm MDN}$ is different from the critical exponent $\beta_{\rm UCN}=1/(3-\gamma)$ for the uncorrelated networks \cite{Cohen02}.
As summarized in Table~\ref{tab:beta}, $\beta_{\rm MDN}=\beta_{\rm UCN}$ for $\gamma>3$, whereas $\beta_{\rm MDN}\neq\beta_{\rm UCN}$ for $2<\gamma<3$.
Thus, the MD structure affects the critical behavior in heavy-tailed scale-free networks ($2<\gamma<3$).

%%%%%%%%%%%%%%%%%%%%%%%%%%%%%%%%%%
\begin{table}[t!]
\begin{center}
\caption{Critical exponent $\beta$ and percolation threshold $p_{\rm c}$ of site percolation on the MD scale-free networks and uncorrelated scale-free networks having an identical degree distribution. This table holds for the bond percolation (see Conclusion \& discussion). Here $z_{1}=\sum_{k}P(k)$ and $z_{2}=\sum_{k}k(k-1)P(k)$.
    }
\begin{tabular}{l|c|c|c||c} \hline
&$2<\gamma<3$&$3<\gamma<4$&$\gamma>4$&$p_{\rm c}$\\\hline
$\beta_{\rm MDN}$&$\frac{\gamma-1}{3-\gamma}$&$\frac{1}{\gamma-3}$&$1$&$\sqrt{\frac{z_{1}^{\rm A}z_{1}^{\rm B}}{z_{2}^{\rm A}z_{2}^{\rm B}}}$\\\hline
$\beta_{\rm UCN}$&$\frac{1}{3-\gamma}$&$\frac{1}{\gamma-3}$&$1$&$\frac{z_{1}}{z_{2}}$\\\hline
\end{tabular}
\label{tab:beta}
\end{center}
\end{table}
%%%%%%%%%%%%%%%%%%%%%%%%%%%%%%%%%%

%%%%%%%%%%%%%%%%%%%%%%%%%%%%%%%%%%
%%%%%%%%%%%%%%%%%%%%%%%%%%%%%%%%%%
\section{Numerical check}\label{sec:numerical}

In this section, we confirm the validity of our theoretical result for $\beta$ as discussed in the previous section.
We can assume that for a network with a finite percolation threshold, $p_{\rm c}>0$, the relative size, $S(N,p)$, of the giant component near the percolation threshold behaves as
\begin{eqnarray}
	S(N,p)=N^{-\beta/\nu_{N}}f(|p-p_{\rm c}|N^{1/\nu_{N}}),
	\label{eq:fss_for_S}
\end{eqnarray}
where $N$ is the system size, $f(z)$ is a scaling function and $\nu_{N}$ is a (volume-based) correlation critical exponent. In a $d$-dimensional system, $\nu_{N}=\nu d$ where $\nu$ is the (length-based) correlation critical exponent related to the correlation length, $\xi$: $\xi\sim|p-p_{\rm c}|^{-\nu}$.
The $N$ dependence of the pseudo percolation threshold, $p_{\rm c}(N)$, is given as
\begin{eqnarray}
	p_{\rm c}(N)-p_{\rm c}= aN^{-1/\nu_{N}},
	\label{eq:nu}
\end{eqnarray}
where $p_{\rm c}=p_{\rm c}(\infty)$ and $a$ is a constant \cite{Stauffer}. 
Substituting Eq.~(\ref{eq:nu}) into Eq.~(\ref{eq:fss_for_S}), we have the $N$ dependence of $S$ at the pseudo percolation threshold as
\begin{eqnarray}
	S(N,p_{\rm c}(N))= bN^{-\beta/\nu_{N}},
	\label{eq:beta/nu}
\end{eqnarray}
where $b=f(a)$.
When $2<\gamma<3$ in which $p_{\rm c}=0$ and the giant component size behaves as Eq.~(\ref{eq:S23}), we expect a finite-size scaling of $S$ as
\begin{eqnarray}
	S(N,p)=N^{-(\beta+1)/\nu_{N}}f(pN^{1/\nu_{N}}),
	\label{eq:fss_for_S2}
\end{eqnarray}
and
\begin{eqnarray}
	S(N,p_{\rm c}(N))/p_{\rm c}(N)= bN^{-\beta/\nu_{N}},
	\label{eq:beta/nu2}
\end{eqnarray}
instead of Eqs.~(\ref{eq:fss_for_S}) and (\ref{eq:beta/nu}) \cite{Radicchi15}.
Fitting  numerical data by Eq.~(\ref{eq:beta/nu2}) for $2<\gamma<3$ [Eq.~(\ref{eq:beta/nu}) for $\gamma>3$] with the help of Eq.~(\ref{eq:nu}), one finds the critical exponents, $\beta$ and $\nu_{N}$.
Thus, we numerically evaluate $\beta$ for the MD networks with a given value of $\gamma$ to validate the relation, $\beta=(\gamma-1)/(3-\gamma)$ [$\beta=1/(\gamma-3)$], theoretically expected for the MD networks with $2<\gamma<3$ (${3<\gamma<4}$).

%%%%%%%%%%%%%%%%%%%%%%%%%%%%%%%%
\begin{figure}[t]
\begin{center}
\includegraphics[width=0.45\textwidth]{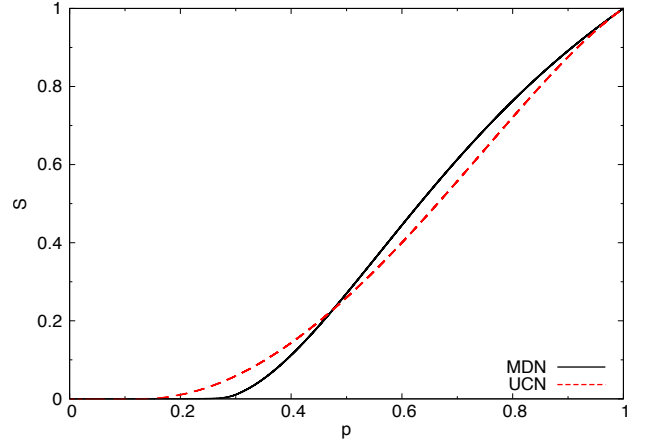}
\caption{
Relative size, $S$, of the largest component as a function of the occupation probability, $p$. 
The solid and dashed lines are the results for the MD networks and uncorrelated networks, respectively. We utilize the scale-free networks with $\gamma=3.5$. The system size is set as $N=1280000$.}
\label{fig:pofS}
\end{center}
\end{figure}
%%%%%%%%%%%%%%%%%%%%%%%%%%%%%%%%

The MD networks employed for our numerical confirmations are generated as follows.
First, according to the Dorogovtsev--Mendes--Samukhin (DMS) model \cite{Dorogo00}, a degree sequence obeying $P_{\rm A}(k)\sim k^{-\gamma}$ with $N_{\rm A}$ nodes and $z_{1}^{\rm A}=6$ is generated \cite{comment1}, which represents group A. 
Second, $N_{\rm B}=3N_{\rm A}$ nodes with $z_{1}^{\rm B}=k_{\rm min}=2$ are prepared as group B. 
Finally, to construct a network, we repeatedly choose a stub at random from each of the groups and join the stubs until all stubs are used up. 
The entire network has $N=4N_{\rm A}$ nodes and the average degree of $z_{1}=3$. 
For comparisons, uncorrelated networks with the same degree distribution (\ref{eq:degree_distribution}) are realized by randomizing the MD networks under preserving the degree of each node. 
The site percolation process is performed numerous times on the MD networks and uncorrelated ones.
Using the Newman--Ziff algorithm \cite{Newman01}, we obtain the average size of the largest component generated by the site percolation process on the networks.

%%%%%%%%%%%%%%%%%%%%%%%
\begin{figure}[tttt]
\begin{center}
\subfigure{
\includegraphics[width=0.45\textwidth]{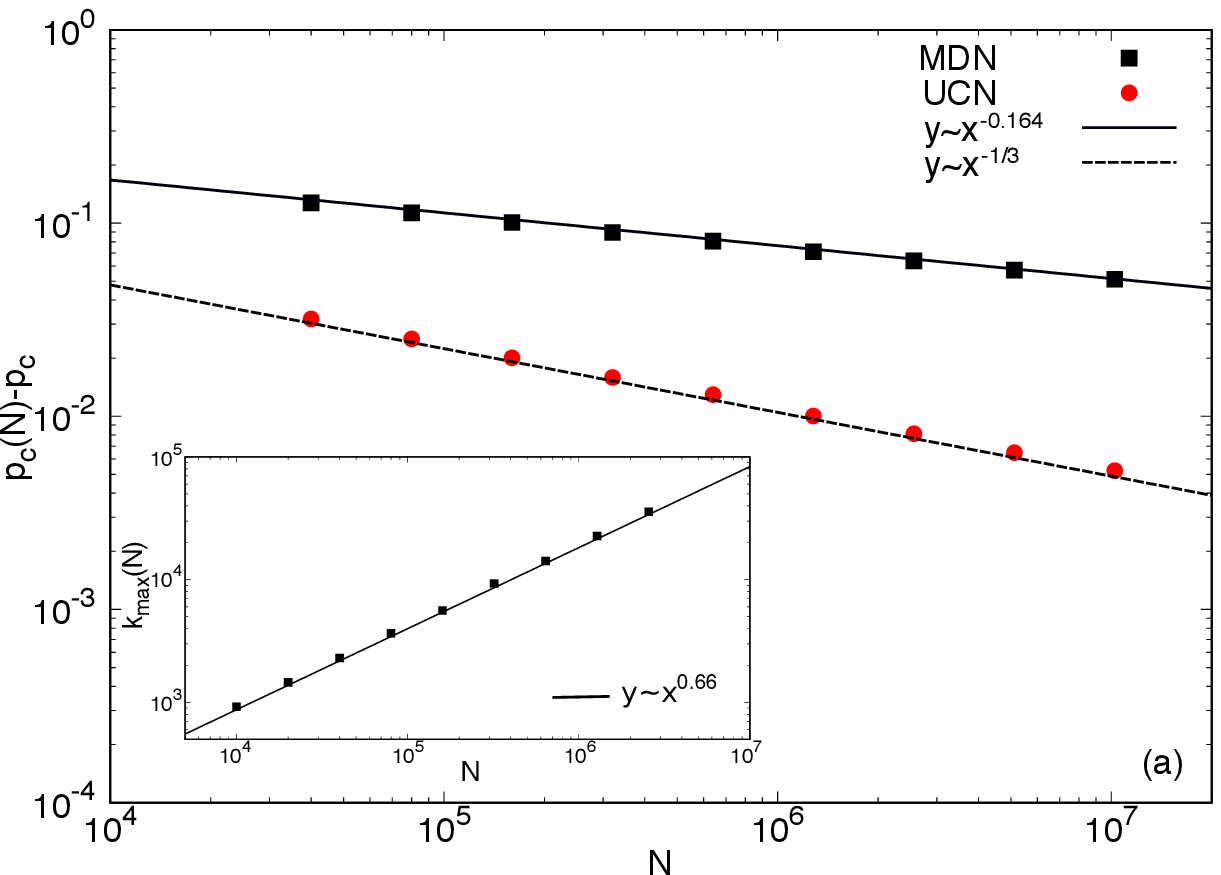}}
	\subfigure{
\includegraphics[width=0.45\textwidth]{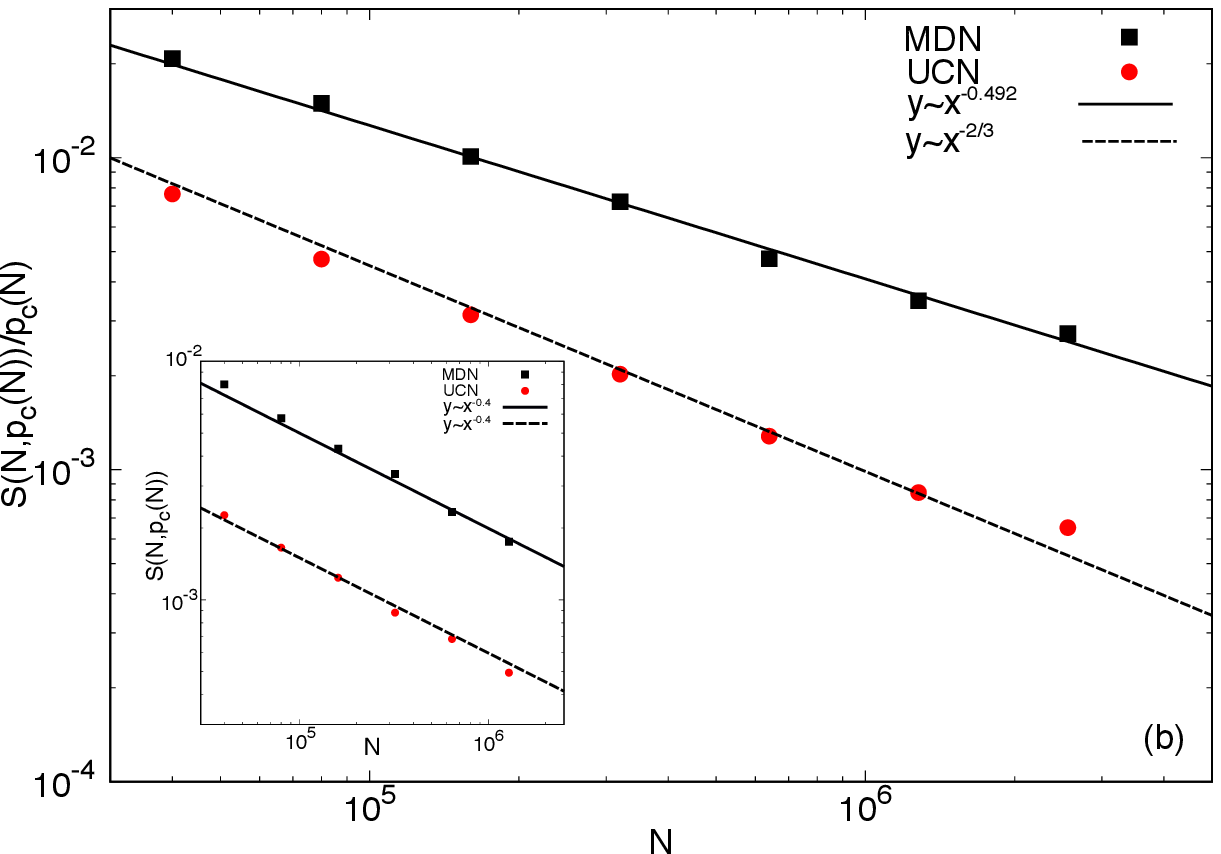}}
\end{center}
\caption{
(a) System size dependence of $p_{\rm c}(N)$.
The black filled squares and red filled circles represent the simulation results for the MD networks and uncorrelated networks, respectively. 
The slopes, $\nu_{N}^{-1}$, of the solid and dashed lines are $-0.164$ and $-1/3$, respectively. 
The system size dependence of the maximum degree, $k_{\rm max}$, for the present networks is displayed in the inset of (a).
Each symbol is an average of over 100 realizations. 
(b) System size dependence of $S(N,p_{\rm c}(N))/p_{\rm c}(N)$ at the pseudo percolation threshold, $p_{\rm c}(N)$. 
The black filled squares and red filled circles represent the simulation results for the MD networks and uncorrelated networks, respectively. 
The slopes, $\beta/\nu_{N}$, of the solid and dashed lines are $-0.51$ and $-2/3$, respectively.
In (b), we generate $30$ network realizations and perform site percolation $30$ times on each realization to take the average of $S(N,p_{\rm c}(N))/p_{\rm c}(N)$.
Except for the inset of (b), we utilize the scale-free networks with $\gamma=2.5$, which are generated by the algorithm explained in the main text. 
In the inset of (b), the system size dependence of $S(N,p_{\rm c}(N))$ for networks with $\gamma=3.5$ is displayed in order to examine Eq.~(\ref{eq:beta/nu}).}
\label{fig:fss}
\end{figure}
%%%%%%%%%%%%%%%%%%%%%%%
%%%%%%%%%%%%%%%%%%%%%%%

First, we examine the relative size, $S$, of the largest component.
Figure~\ref{fig:pofS} shows $S_{\rm MDN}$ (the solid line) for the MD networks with degree exponent $\gamma=3.5$ and $S_{\rm UCN}$ (the dashed line) for the corresponding uncorrelated networks. 
The percolation threshold, $p_{\rm c} $, for the MD networks is larger than that for the uncorrelated ones, as shown in Table~\ref{tab:beta}.
Next, by numerically evaluating the critical exponents, $\nu_{N}$ and $\beta$, for the uncorrelated scale-free networks, we confirm the validity of the present scalings (\ref{eq:nu}) and (\ref{eq:beta/nu2}).
Figure~\ref{fig:fss}(a) shows the $N$ dependence of the pseudo percolation threshold, $p_{\rm c}(N)$, for the uncorrelated scale-free networks with $\gamma=2.5$. 
The red filled circles represent the simulation result.
Here, the pseudo percolation threshold, $p_{\rm c}(N)$, for the uncorrelated networks is estimated by substituting the numerically obtained degree distributions for the Molly--Reed criterion, $p_{\rm c}(N)=z_{1}/z_{2}$, where $z_{2}=\sum_{k}k(k-1)P(k)$ \cite{Newman01,Cohen00}.
According to \cite{Cohen02}, we theoretically obtain $\nu_{N}^{\rm UCN}$ as $\nu_{N}^{\rm UCN} =(\gamma-1)/(3-\gamma)$ when the maximum degree $k_{\rm max}$ behaves as $k_{\rm max} \sim N^{1/(\gamma-1)}$ \cite{Dorogo02} [see the inset of Fig.~3(a)].
The dashed line is drawn by using the theoretical value, $(\nu_{N}^{\rm UCN})^{-1}=1/3$ for $\gamma=2.5$.
The red filled circles lie on the dashed line, which means that Eq.~(\ref{eq:nu}) for the uncorrelated networks is correct.
Figure~\ref{fig:fss}(b) shows the $N$ dependence of the largest component size over the pseudo percolation threshold, $S(N,p_{\rm c}(N))/p_{\rm c}(N)$. 
The red filled circles representing the simulation results lie on the dashed line drawn using Eq.~(\ref{eq:beta/nu2}) with the theoretical values of $\beta_{\rm UCN}=1/(3-\gamma)=2$ and $(\nu_{N}^{\rm UCN})^{-1}=1/3$.
Thus the finite-size scalings for $S$, Eqs.~(\ref{eq:nu}) and (\ref{eq:beta/nu2}), succeed in capturing the critical behavior of percolation on the uncorrelated networks, as was reported in \cite{Radicchi15}. 

Finally, we apply finite-size scaling analysis to the MD networks.
The black filled squares in Fig.~\ref{fig:fss}(a) represent the simulation results for the MD scale-free networks with $\gamma=2.5$.
The slope of the black filled squares is estimated as $\nu_{N}^{-1}=0.164$ and is different from that for the uncorrelated networks (the red filled circles), which indicates that the percolation critical exponents of the MD networks differs from those of the uncorrelated ones. 
Here, the solid line is a guide to the eye with a slope of $-0.164$.
Substituting $\beta=(\gamma-1)/(3-\gamma)$ and $\nu_{N}^{-1}=0.164$ for $\beta/\nu_{N}$, we have $\beta/\nu_{N}=0.492$ for $\gamma=2.5$.
In Fig.~\ref{fig:fss}(b), we depict a solid line with the slope, $\beta/\nu_{N}=0.492$. 
The line is parallel to the black filled squares (simulation results for $\gamma=2.5$), which supports the theoretical result for $2<\gamma<3$. 
On the one hand, in the inset of Fig.~\ref{fig:fss}(b), we examine Eq.~(\ref{eq:beta/nu}) for $\gamma=3.5$. Both slopes of plots for the MD networks and uncorrelated ones are parallel. This means that the MD networks with $\gamma=3.5$ and corresponding uncorrelated networks have the same value of $\beta$. 
All results exhibit the validity of our theoretical arguments.

%%%%%%%%%%%%%%%%%%%%%%%%%%%%%%%%%%
%%%%%%%%%%%%%%%%%%%%%%%%%%%%%%%%%%
\section{Conclusion \& Discussion}\label{sec:conclusion}

We have studied the site percolation on a maximally disassortative (MD) network, which has a maximally negative degree-degree correlation and is realized as a bipartite network.
Based on the generating function method for bipartite networks, we have clarified the percolation threshold and the order parameter critical exponent of the MD networks.
We have found that the critical behavior of site percolation on the MD networks is different from that on  uncorrelated networks when networks are heavy-tailed so that $P(k)\sim k^{-\gamma}$ with $2<\gamma<3$. The MD networks with $\gamma>3$ have the same critical behavior as uncorrelated networks with the identical degree distribution. 
These results have been numerically confirmed by a finite-size scaling analysis.

We should mention how the {\it bond} percolation in which each edge is retained with probability $p$ and removed otherwise behaves in our MD network because the site-bond percolation universality in uncorrelated scale-free networks breaks in terms of the scaling for the giant component size, $S$, when the percolation threshold is zero \cite{Radicchi15}.
Our calculation can be applied to the bond percolation on the MD networks.
For the bond percolation problem, it is unchanged that $H_1^{\rm A}(x)$ and $H_1^{\rm B}(x)$ follow Eqs.(\ref{eq:H1A}) and (\ref{eq:H1B}), respectively, meaning that $p_c$ and $\beta$ for bond percolation take the same value shown in Table~\ref{tab:beta} because $p_c$ and $\beta$ are determined by Eqs.~(\ref{eq:H1A}) and (\ref{eq:H1B}).
Thus, the site-bond percolation universality coincides in terms of the order parameter, $\epsilon$.
On the one hand, the generating functions (\ref{eq:H0A}) and (\ref{eq:H0B}) are replaced by $H_{0}^{\rm A(B)}(x)=xG_{0}^{\rm A(B)}[H_{1}^{\rm B(A)}(x)]$.
Expanding $H_{0}^{\rm A(B)}(1)=G_{0}^{\rm A(B)}[u(v)]$, we find $S[= 1-H_{0}^{\rm tot}(1)]\sim \epsilon$ for the bond percolation (whereas $S\sim p\epsilon$ for the site percolation).
The bond percolation on the MD networks shows $S\sim p^{\beta_{\rm MDN}}$ for $2<\gamma<3$ ($p_{\rm c}=0$), which is different from the case of the site percolation [Eq.~(\ref{eq:S23})].
Therefore, the site-bond percolation universality breaks in terms of the singularity of $S$, as was reported in \cite{Radicchi15}.
It can be confirmed numerically by means of the finite-size scaling argument for Monte-Carlo data (not shown).

In this work, we have focused on the MD networks whose disassortativity is realized by setting $P_{\rm B}(k)=\delta_{k2}$.
Here, let us consider the case that $P_{\rm B}(k)$ has a large but finite second moment to test the robustness of the result for the MD networks with the degree exponent $2<\gamma<3$ [the second moment of $P_{\rm A}(k)$ diverges].
Such a network is disassortative in that the covariance of the degree-degree joint probability is infinitely negative.
In this situation, Eq.~(\ref{eq:v}) is changed to $v=1-p+p\sum kP_{\rm B}(k)u^{k-1}/z_{1}^{\rm B}\sim 1-z_{2}^{\rm B}p\epsilon/z_{1}^{\rm B}$.
A similar calculation for the summation in Eq.~(\ref{eq:ep}) yields the same exponent displayed in Table~\ref{tab:beta}. The critical property is unchanged as long as $P_{\rm B}(k)$ has a finite second moment.
A work concerning a biased bond percolation \cite{Hooyberghs10} may be useful for considering the case that the second moment of $P_{\rm B}(k)$ diverges. 
In the biased bond percolation of parameter $\alpha$, each edge connecting a degree-$k$ node and a degree-$k'$ node is retained with a probability proportional to $(kk')^{-\alpha}$ \cite{Hooyberghs10,Hooyberghs10-2}.
The biased bond percolation reduces to ordinary bond percolation when $\alpha\to 0$.
In \cite{Hooyberghs10}, Hooyberghs \textit{et al.} considered the biased bond percolation on bipartite networks whose both groups obey a power-law degree distribution, i.e., $P_{\rm A(B)}(k)\sim k^{-\gamma_{\rm A(B)}}$ ($\gamma_{\rm A}<\gamma_{\rm B}$) and showed that for $\alpha=0$ and a fixed value of $\gamma_{\rm B}>3$,
the order parameter critical exponent $\beta$ takes the same value in Table~\ref{tab:beta}.
In addition, for the case of $2<\gamma_{\rm A}<\gamma_{\rm B}<3$, they has found the critical behavior depending on two degree exponents.
While it may be hard to argue the degree correlation of networks with $2<\gamma_{\rm A}<\gamma_{\rm B}<3$, we expect that they are strongly disassortative.

Finally, we have focused on networks having only a nearest neighbor degree correlation.
Real-world networks, however, have a long-range degree correlation, which cannot be captured by any nearest neighbor degree correlation \cite{Fujiki17,Rybski10,Fujiki18,Mayo15,Fujiki19}.
Little is known about what long-range correlated structures induce in the percolation transition.
Some numerical studies \cite{Noh07,Menche11,Valdez11} on correlated networks with the tunable degree-degree correlation have suggested that an unusual type of phase transition originates from something beyond the nearest neighbor degree correlation.
Further studies to understand how correlated structures beyond the nearest neighbor degree correlations affect the critical phenomena of the networks should be conducted.

\acknowledgments
S.M.\ and T.H.\ acknowledge the financial support from JSPS (Japan) KAKENHI Grant Number JP18KT0059.
S.M.\ was supported by a Grant-in-Aid for Early-Career Scientists (No.~18K13473) and Grant-in-Aid for JSPS Research Fellow (No.~18J00527) from the Japan Society for the Promotion of Science (JSPS) for performing this work.
T.H.\ acknowledges the financial support from JSPS (Japan) KAKENHI Grant Numbers JP16H03939 and JP19K03648.

\section{Appendix}
We review a treatment of $G_{1}^{\rm A}(v)=\sum_{k}kP_{\rm A}v^{k-1}/z_{1}^{\rm A}$ briefly.
For simplicity, let us consider $P_{\rm A}(k)\sim ck^{-\gamma}$ with $2<\gamma<3$.
From the normalization condition, $G_{1}^{\rm A}(v)=1$ for $v=1$.
However, its derivative,
\begin{eqnarray}
	G_{1}^{\rm A'}(v)=\sum_{k}\frac{ck(k-1)k^{-\gamma}}{z_{1}^{\rm A}}v^{k-2}
\end{eqnarray}
diverges in the limit $v\to 1$.
Because $G_{1}^{\rm A'}(v)$ converges for $0\le v<1$, $G_{1}^{\rm A'}(v)$ has the asymptotic form
\begin{eqnarray}
	G_{1}^{\rm A'}(v)\sim \frac{c}{z_{1}^{\rm A}}\Gamma(3-\gamma)(1-v)^{\gamma-3},
	\label{eq:tauber}
\end{eqnarray}
in $v\to 1-$.
Here we use a Tauberian theorem (see e.g.\ Chap.~XIII in \cite{Feller68}).
From the integral of Eq.~(\ref{eq:tauber}),
\begin{eqnarray}
	G_{1}^{\rm A}(1)-G_{1}^{\rm A}(v)\sim \frac{c}{z_{1}^{\rm A}}\Gamma(3-\gamma)\int_{v}^{1} (1-v')^{\gamma-3}dv',
\end{eqnarray}
we have
\begin{eqnarray}
	G_{1}^{\rm A}(v)\sim G_{1}^{\rm A}(1)+\frac{c}{z_{1}^{\rm A}}\Gamma(2-\gamma)(1-v)^{\gamma-2},
	\label{eq:tauber2}
\end{eqnarray}
which is equivalent to Eq.~(\ref{eq:exp23}).
In a similar way, we can derive Eq.~(\ref{eq:expansion}).

\bibliography{ref}

\begin{thebibliography}{10}
\bibitem{comment1}
The degree exponent, $\gamma$, in the DMS model is given by $\gamma=3+a/m$,
  where $a$ is the predetermined attractiveness of each node and $m$ is the
  number of stubs of each adding node. For the generated networks in this
  study, we set a clique consisting of seven nodes as the initial condition and
  $m=3$.
\end{thebibliography}

\end{document}